\documentclass[onecolumn,aps,floatfix,pra,superscriptaddress,nofootinbib]{revtex4-2}

\usepackage{color}
\usepackage{graphicx}
\usepackage{amsmath}%
\usepackage{amssymb}%
\usepackage{mathrsfs}
\usepackage{enumerate}
\usepackage{upgreek}
\usepackage{mathtools}
\usepackage[subnum]{cases}
\usepackage{dsfont}

\newcommand{\tr}{ \text{tr} }

\newcommand{\ti}{ \tilde }

\newcommand{\hb}{ \hbar }
\newcommand{\si}{ \sigma }
\newcommand{\om}{ \omega }
\newcommand{\Om}{ \Omega }
\newcommand{\ga}{ \gamma }
\newcommand{\la}{ \langle }
\newcommand{\ra}{ \rangle }

\newcommand{\da}{ \dagger }
\newcommand{\Del}{ \Delta }
\newcommand{\lam}{ \lambda }
\newcommand{\re}{ \text{Re} }
\newcommand{\im}{ \text{Im} }

\begin{document}

\title{Quantum correlations under environmental decoherence}

\author{S. V. Mousavi}
\email{vmousavi@qom.ac.ir}
\affiliation{Department of Physics, University of Qom, Ghadir Blvd., Qom 371614-6611, Iran}

\begin{abstract}

Taking a system of two coupled qubits described by a X-shaped state and interacting through an anisotropic Heisenberg XY interaction, we examine the evolution of quantum entanglement and a few quantum correlations beyond entanglement, local quantum uncertainty and measurement-induced nonlocality, under the environmental decoherence both for zero and finite temperatures. The relation between concurrence and log negativity as two well-known quantifiers of entanglement is argued. It will be proven that measurement-induced nonlocality equals correlated coherence.
The interaction of qubits with the environment causes quantum entanglement to suddenly die for independent qubits, but other correlations do not experience this phenomenon. The time of entanglement sudden death is calculated analytically for zero temperature, while numerically for finite temperatures.  
Contrary to its usual destructive role, the environment plays a constructive role in some situations, inducing quantum correlations even when the initial quantum correlations are zero.
The steady state quantum correlations, being independent of the initial state, are found to remain all non-zero for low, finite temperatures. It is found that quantum correlations beyond entanglement are more robust against temperature than entanglement.
The zero-temperature steady state exhibits less local quantum uncertainty than the other correlations.

\end{abstract}

\maketitle


\section{Introduction}\label{sec1}

Quantum states are fundamentally different from classical states. Any measurements made on one part of a bipartite state will make the results uncertain. The integral nature of quantum states is the reason for this randomness, not the measuring device. 
%
Only states, which can be described by a classical probability distribution, are invariant under such local measurements. 
Perhaps the most striking feature of quantum mechanics is the quantum entanglement (QE). It is at the heart of quantum information processing, being a resource for doing tasks like quantum computing, quantum teleportation and quantum key distribution \cite{NiCh-book-2000}. Nonetheless, in the early age of quantum information science, some researchers questioned entanglement as the source of power for quantum computation \cite{MiLaSaKn-PRA-2002}.
QE is not the only correlation which exists between different parts of a multi-party quantum system. 
Although pure states can be either uncorrelated or just entangled \cite{AdBrCi-JPA-2016}, it has been proven that certain quantum {\it separable mixed} states exhibit quantum correlations, the so-called quantum discord \cite{OlZu-PRL-2001, HeVe-JPA-2001}.
In addition to the significant role of quantum correlations in efficient quantum communication and computational tasks, they are also important in figuring out cooperative phenomena in multi-party quantum systems.
Among various beyond entanglement quantum correlations have already been identified, one can mention local quantum uncertainty (LQU) \cite{GiTuAd-PRL-2013} and measurement-induced nonlocality (MIN)  \cite{LuFu-PRL-2011, HuFa-NJP-2015}. While the former is introduced as the minimum skew information \cite{WiYa-PNAS-1963} achievable on a single local measurement, the latter is interpreted as the maximum global effect caused by locally invariant measurements. To see some applications of quantumness beyond entanglement see \cite{AdBrCi-JPA-2016}.

Beside quantum correlations, quantum coherence \cite{BaPl-PRL-2014}, being cornerstone of quantum computation and quantum communication theory, has a fundamental and practical significance for developing the quantum mechanics \cite{Huetal-PR-2018}. Contrary to quantum correlations, being defined for systems with at least two parties, quantum coherence measures can also be defined for single systems. However, both arise from quantum superposition principle, thus being interrelated. In this regard, the link between quantum coherence and entanglement have already been identified \cite{StSiDhBeAd-PRL-2015, QiGaYa-JPA-2017}. Furthermore, the concept of ``correlated coherence"(CC) \cite{Taetal-PRA-2016} has been introduced and its connection to quantum discord and entanglement has been argued. The idea is splitting the coherence contained in the global state into two portions, those stored locally in the marginal (reduced) states and that stored non-locally only in the correlated state, and then subtracting local portions from the correlated one. For X states, we'll find that MIN equals CC.

Since a quantum system cannot be completely isolated from its environment, its influence on the system's dynamics must be taken into account. 
Two approaches have been developed to study dynamics of open quantum systems: the quantum operations formalism and the approach of master equations, which each one has its own place \cite{NiCh-book-2000}. 
The latter strategy employs a differential equation, which, in addition to the conventional unitary term for an isolated system, includes a non-unitary term responsible for interaction with the surrounding environment. 
In the decoherence process, one of the most striking results is the complete loss of entanglement in a finite time, the so-called ``entanglement sudden death" (ESD) implying complete termination of entanglement after a finite time interval without a long-lasting smooth decrease \cite{YuEb-PRL-2004, YuEb-PRL-2006, EbYu-Sci-2007}. Mathematically, ESD is based on the cutoff in the definition of concurrence, but it is difficult to find a physical interpretation. However, its connection to energy transfer between the system and its surroundings has been discussed \cite{CuLiYi-PLA-2007}.
On contrary, the environment-induced \cite{FiTa-PRA-2006} or  decoherence-induced entanglement \cite{DaLi-JPB-2007} has been worked out. Here, environment plays a constructive role, in contrast to its customary destructive role, inducing QE. Furthermore, entanglement re-birth or revival has also been reported. 
As quantum correlations and particularly quantum entanglement form the basis of quantum technologies, a significant concern pertains to the prevention or management of disentanglement resulting from decoherence in realistic physical systems. Different approaches have been proposed for retarding or avoiding ESD \cite{ZhMaXiGu-EPJD-2010, SaBj-PRA-2008, ChSa-EPJD-2018, NaAr-PS-2024}.

In this work, we examine a system of two localized coupled qubits that are in interaction with an external magnetic field. The interaction with the environment is introduced by a non-unitary term in the Lindblad equation. We are going to solve this equation analytically first for a number of initial states in the form X and then for a general X-shaped state; and study the evolution of different quantum correlations for different emission rates as well as different interaction intensities. It is imperative to determine whether quantum correlations beyond entanglement experience sudden death or not. Furthermore, it is important to conduct an investigation into the phenomenon of correlation induced by the environment. After all this, it will be time to consider the effect of temperature on quantum correlations. How does this parameter affect ESD?

It is noteworthy that X states have been extensively analyzed in the literature from different aspects. Analytical expressions for diverse measures of quantum correlations have been derived for general X states \cite{YuEb-QIC-2007, Mi-IJTP-2012, KhDaSa-PS-2019, YuHa-PLA-2023} and their dynamical behavior has been examined. For instance, see \cite{ChZhYuYiOh-PRA-2011, SlDaAh-QIP-2019, SlDaAh-QIP-2018, TiGuChWa-QIP-2023, ChBhPa-EPJP-2022, KhErDa-EPJP-2021}.
Furthermore, it is worth mentioning that investigation of the effect of decoherence on quantum correlations through other approaches like {\it intrinsic} decoherence in the Milburn framework \cite{Mi-PRA-1991, MoMi-EPJP-2024} or quantum channels \cite{NiCh-book-2000} had been the subject of many studies \cite{ChHaNa-QIP-2021, MuCh-QIP-2021, BeAnDa-EPJP-2022, Zi-IJTP-2016, Guetal-QIP-2019, PoHaNaKuSh-AEJ-2023, MuSa-QIP-2018}.

The paper is organized as follows. In Section \ref{sec: QC} we concisely address quantifiers of the mentioned quantum correlations and also the quantum coherence. In Section \ref{sec: model} we introduce the model and solve the corresponding Lindblad equation for some X-shaped states.
Section \ref{sec: temp} includes steady state solution of the Lindblad equation in a finite temperature for a general X state.  
Section \ref{sec: results} is devoted to computations, both analytical and numerical, and discussion of the results. Finally, in Section \ref{sec: coclusion} some concluding results are presented.

\section{Some Quantum Correlations; entanglement and beyond} \label{sec: QC}

In this section we briefly review QE, LQU and MIN. At the end, we consider CC and relate it to MIN.
%
%
Different measures have been introduced for quantification of QE. One of the most famous measures of entanglement is the concurrence \cite{HiWo-PRL-1997, Wo-PRL-1998} which for a two-qubit state $ \rho $ is defined as 
\begin{eqnarray} \label{eq: con-def}
 \mathcal{C}(\rho) &=& \max \{ 0, \lam_1 - \lam_2 - \lam_3 - \lam_4 \},
\end{eqnarray}
$ \lam_i $s  being the eigenvalues of the Hermitian matrix $ \sqrt{ \sqrt{\rho} \ti{\rho} \sqrt{\rho} } $ in non-increasing order. Here, $ \ti{\rho} $ is the spin-flipped Hermitian density matrix given by $ (\si_y \otimes \si_y) \rho^* (\si_y \otimes \si_y) $; $ \si_y $ being the second Pauli matrix and $ \rho^* $ the complex conjugate of $ \rho $ which due to the Hermiticity of $\rho$ equals its transpose, $ \rho^* = \rho^T  $. 
Two matrices $ S = \sqrt{\rho} \ti{\rho} \sqrt{\rho} $ and $ R = \rho \ti{\rho} $ are connected through the invertible matrix $ \sqrt{\rho} $ i.e., $ R = (\sqrt{\rho}) S (\sqrt{\rho})^{-1} $. Thus, they are similar, having the same spectrum. In this way, one sees $ \lam_i $s are the square roots of the eigenvalues of $S$ and thus the non-Hermitian matrix $R$. 
Note that since $ \si_y \otimes \si_y $ is its own inverse, $ (\si_y \otimes \si_y)^{-1} = \si_y \otimes \si_y $ then $ \ti{\rho} $ is similar to $ \rho^* $ which its eigenvalues are just those of $ \rho $.
Therefore, both $ \rho^* $ and its similar matrix $ \ti{\rho} $ are non-negative. Now, since both $\rho$ and $\ti{\rho}$ are non-negative, eigenvalues of their product $ R = \rho \ti{\rho} $ are non-negative \cite{Bh-book-1997}\footnote{See page 256.}.
Concurrence, defined in this way, becomes zero for separated states but takes the value one for maximally entangled states in $ \mathbb{C}^2 \otimes \mathbb{C}^2 $ systems \cite{Dhetal-book-2017}. It has been argued that the concurrence is related to the entanglement of formation, which itself is an entropic entanglement monotone \cite{HiWo-PRL-1997, Wo-PRL-1998}.
In the basis $ \{ |00\ra, |01\ra, |10\ra, |11\ra \} $, a X-state has the form
%
\begin{eqnarray} \label{eq: X-shaped}
\rho &=& 
\begin{pmatrix}
\rho_{11} & 0 & 0 & \rho_{14} \\
0 & \rho_{22} & \rho_{23} & 0 \\
0 & \rho_{32} & \rho_{33} & 0 \\
\rho_{41} & 0 & 0 & \rho_{44} 
\end{pmatrix}
\end{eqnarray}
with constraints
\begin{eqnarray} \label{eq: X-constraints}
\qquad \sum_{i=1}^4 \rho_{ii} = 1, \quad \rho_{22} \rho_{33}  \geq | \rho_{23} |^2 , \quad \rho_{11} \rho_{44}  \geq | \rho_{14} |^2,
\end{eqnarray}
coming from properties of the density operator i.e., unit trace, Hermiticity and positivity. 
The concurrence \eqref{eq: con-def} for the X-state \eqref{eq: X-shaped} is then \cite{Waetal-JPB-2006}
\begin{eqnarray} \label{eq: concur}
\mathcal{C} &=& \max \{ 0, \mathcal{C}_1, \mathcal{C}_2 \}
\end{eqnarray}
where,
\begin{numcases}~
\mathcal{C}_1 = 2 ( |\rho_{14}| - \sqrt{ \rho_{22} \rho_{33} } ) \label{eq: C1}  \\
\mathcal{C}_2 = 2 ( |\rho_{32}| - \sqrt{ \rho_{11} \rho_{44} } ) \label{eq: C2}
\end{numcases}
As these equations show, if and only if either $ \rho_{22} \rho_{33}  < | \rho_{14} |^2 $ or $ \rho_{11} \rho_{44}  < | \rho_{23} |^2 $ then the state \eqref{eq: X-shaped} is entangled. Both conditions cannot be fulfilled simultaneously \cite{AlRaAl-PRA-2010}.
Our model which will be considered in section \ref{sec: model} can be regarded as two effective two-level atoms that are subject to decoherence.
As such, $ \mathcal{C}_1 $ provides a measure of an entanglement produced by the two-photon coherence that is between the ground state, with no photons, and the two-photon excited state ($|11\ra \leftrightarrow |00\ra $), whereas $ \mathcal{C}_2 $ provides a measure of an entanglement produced by the one-photon coherence corresponding to the coherence that is present between two different one-photon states ($|10\ra \leftrightarrow |01\ra$).
If an entanglement is generated among the one-photon states, the magnitude of the entanglement is restricted by the populations of the no-photon and two-photon states, and conversely, if an entanglement is created among the two-photon states, the magnitude of the entanglement is restricted by the population of the one-photon states $\rho_{22}$ and $ \rho_{33} $ \cite{Fi-FPC-2010}.

The basis of collective states of the system, or the Dicke states, which are defined as
\begin{numcases}~ 
| e \ra = | 0 0 \ra \label{eq: ee} \\
| g \ra = | 1 1 \ra \label{eq: gg} \\
| s \ra = ( | 0 1 \ra + | 1 0 \ra ) / \sqrt{2} \label{eq: symmetric} \\
| a \ra = ( | 0 1 \ra - | 1 0 \ra ) / \sqrt{2} \label{eq: anti-symmetric} 
\end{numcases}
is a particularly useful choice for working with. Here, the excited and ground states $|e\ra$ and $|g\ra$ are separable, while the symmetric and antisymmetric states $|s\ra$ and $|a\ra$ are maximally entangled. The X-state \eqref{eq: X-shaped} in this new basis is then expressed as,
\begin{eqnarray} \label{eq: rho_Dicke}
\varrho &=&
\left(
\begin{array}{cccc}
 \rho_{11} & \rho_{14} & 0 & 0 \\
 \rho_{41} & \rho_{44} & 0 & 0 \\
 0 & 0 & \frac{1}{2} (\rho_{22}+\rho_{33}) + \re\{\rho_{32}\} & \frac{1}{2} (\rho_{22} - \rho_{33}) + i~ \im\{\rho_{32}\} \\
 0 & 0 & \frac{1}{2} (\rho_{22} - \rho_{33}) - i~ \im\{\rho_{32}\} & \frac{1}{2} (\rho_{22}+\rho_{33}) - \re\{\rho_{32}\} \\
\end{array}
\right)
\equiv
\left(
\begin{array}{cccc}
 \varrho_{ee} & \varrho_{eg} & 0 & 0 \\
 \varrho_{ge} & \varrho_{gg} & 0 & 0 \\
 0 & 0 & \varrho_{ss} & \varrho_{sa} \\
 0 & 0 & \varrho_{as} & \varrho_{aa} \\
\end{array}
\right)
\end{eqnarray}
which has the block-diagonal form. Note that the diagonal elements are interpreted as populations, while the off-diagonal ones represent coherences \cite{FiTa-JCMSE-2010}.
In terms of the density matrix in the Dicke basis, $\mathcal{C}_1$ and $\mathcal{C}_2$ of Eq. \eqref{eq: concur} is given by \cite{FiTa-JCMSE-2010, NuMiPa-MPLA-2023}
\begin{eqnarray} 
\mathcal{C}_1 &=& 
2 \left\{ |\varrho_{eg}| - \left[ \left( \frac{ \varrho_{ss} + \varrho_{aa} }{2} \right)^2 - ( \re\{ \varrho_{sa} \} )^2  \right]^{1/2} \right\}  \label{eq: C1-Dicke}  \\
\mathcal{C}_2 &=& 
2 \left\{ \left[ \left( \frac{ \varrho_{ss} - \varrho_{aa} }{2} \right)^2 + ( \im\{ \varrho_{sa} \} )^2  \right]^{1/2} - 
\sqrt{ \varrho_{ee} \varrho_{gg} } \right\} \label{eq: C2-Dicke}
\end{eqnarray}
%


As another natural and widely used entanglement measure, negativity for a bipartite state $ \rho $ is defined through the trace norm of the partially transposed density matrix \cite{ViWe-PRA-2002}, 
\begin{eqnarray} \label{eq: neg}
\mathcal{N} &=& \frac{ || \rho^{T_B} ||_1 - 1  }{2},
\end{eqnarray}
where $ || X ||_1 = \tr( \sqrt{X^{\da} X} ) $ is the trace norm and $ \rho^{T_B} $ denotes the partial transpose of the density matrix with respect to the party $B$,
\begin{eqnarray} \label{eq: partr-B}
\rho^{T_B} &=& \sum_{i,j} \sum_{k,l} \rho_{ij, kl} | i \ra \la j | \otimes | l \ra \la k |,
\end{eqnarray}
where $ \{ | i \ra \} $ and $ \{ | k \ra \} $ are given local orthonormal basis for subsystems $A$ and $B$ respectively.
It is notable that the spectrum of the partial transposition of the density matrix is independent both of the choice of the local basis and also whether the partial transposition is performed over the party $A$ or $B$ \cite{PlVi-bookchap-2014}.
One can easily check that $ \rho^{T_B} $ is Hermitian
\begin{eqnarray*}
\la i,k | (\rho^{T_B})^{\da} | j, l \ra = \la j, l | \rho^{T_B} | i , k \ra^*
= \la j, k | \rho | i , l \ra^* = \la i, l | \rho | j , k \ra = \la i, k | \rho^{T_B} | j , l \ra ,
\end{eqnarray*}
where in the third equality we have used Hermiticity of the density operator $\rho$.
Now, from Hermiticity of $ \rho^{T_B} $ and the fact that trace is independent of basis, one has that  
\begin{eqnarray} \label{eq: rhoTB-trnorm}
|| \rho^{T_B} ||_1 &=& \tr \bigg( \sqrt{ ( \rho^{T_B} )^2 } \bigg) = \sum_i \sqrt{\lam_i^2}
= \sum_i | \lam_i |,
\end{eqnarray}
$ \lam_i $ being eigenvalues of $\rho^{T_B}$. From Eqs. \eqref{eq: neg} and \eqref{eq: rhoTB-trnorm}, and the fact that $ \sum_i \lam_i = \tr( \rho^{T_B} )  = \tr( \rho ) = 1 $ one obtains
\begin{eqnarray} \label{eq: neg2}
\mathcal{N} &=& \sum_i \frac{ | \lam_i | - \lam_i }{2} = - \sum_{\lam_i < 0}  \lam_i =
\begin{cases}
0 & \rho^{T_B} \text{~is positive definite} \\
- \lam_{\min} & \text{only one negative eigenvalue}
\end{cases}
\nonumber \\
& = & \max \{ 0, - \lam_{\min} \},
\end{eqnarray}
%
where in the third equality we have used the fact that for entangled two qubit states the partiality transposed state $ \rho^{T_B} $ has only one negative eigenvalue \cite{MiGr-JPB-2004} which has been denoted by $ \lam_{\min} $ in Eq. \eqref{eq: neg2} as it is the minimum eigenvalue.
Some authors have defined negativity of two-qubit states as 
\begin{eqnarray} \label{eq: neg-prime}
\mathcal{N}' &=& 2 \mathcal{N} = \max \{ 0, - 2 \lam_{\min} \} 
\end{eqnarray}
which for maximally entangled states where the spectrum of $\lam$ is $ \{ - \frac{1}{2}, \frac{1}{2}, \frac{1}{2}, \frac{1}{2} \} $, takes the largest value 1 \cite{ZyHoSaLe-PRA-1998, VeAuDeMo-JPA-2001}.
For a mixed state of two qubits, the subject of our study, it has been proved that the negativity $ \mathcal{N}' $ of a state can never exceed its concurrence and is always greater than $ \sqrt{ (1-\mathcal{C})^2 + \mathcal{C}^2 } - ( 1 - \mathcal{C} ) $; $ \mathcal{C} $ being the concurrence of the state \cite{VeAuDeMo-JPA-2001}. Now, from Eqs. \eqref{eq: neg2} and \eqref{eq: neg-prime} we have that 
$ \mathcal{N} = \mathcal{N}' / 2 $ and thus from the results of \cite{VeAuDeMo-JPA-2001} we have that 
\begin{eqnarray} \label{eq: concur-neg}
\frac{ \sqrt{ [1-\mathcal{C}]^2 + \mathcal{C}^2(t) } - ( 1 - \mathcal{C} ) }{2} \leq \mathcal{N} \leq \frac{ \mathcal{C} }{2}.
\end{eqnarray}

The logarithmic negativity (log-negativity) as an appropriate quantifier of QE in a bipartite quantum system is defined as \cite{ViWe-PRA-2002, Pl-PRL-2005}
\begin{eqnarray} \label{eq: logneg}
\mathcal{L_N} &=& \log_2( || \rho^{T_B} ||_1 ) = \log_2( 2 \mathcal{N} + 1 ),
\end{eqnarray}
where in the second equality we have used \eqref{eq: neg}.
From Eqs. \eqref{eq: concur-neg} and \eqref{eq: logneg}, and since function $ \log(x) $ is an increasing function of its argument, we get
\begin{eqnarray} \label{eq: concur-logneg}
\log_2 \bigg( \sqrt{ [1-\mathcal{C}]^2 + \mathcal{C}^2 } + \mathcal{C} \bigg) \leq \mathcal{L_N} \leq \log_2 [ \mathcal{C} + 1 ] .
\end{eqnarray}
%
%
%
%

%
%

Now we consider quantum correlations beyond QE. Based of the notion of skew information, authors of \cite{GiTuAd-PRL-2013} have proposed a discord-like scheme for characterizing non-classical correlation via LQU. In this way, LQU has been introduced as the minimum skew information achievable on a single local measurement on the subsystem $A$ and it has been obtained as
\begin{eqnarray} \label{eq: LQU}
\mathcal{U} &=& 1 - \lam_{\max}\{ W \} ,
\end{eqnarray}
where $ \lam_{\max}\{ W \} $ shows the maximum eigenvalue of the matrix
\begin{eqnarray} \label{eq: W-mat}
W_{ij} &=& \tr( \sqrt{\rho} ~ \si^{(A)}_i \otimes \mathds{1}^{(B)} ~ \sqrt{\rho(t)} ~ \si^{(A)}_j \otimes \mathds{1}^{(B)} ) .
\end{eqnarray}
Due to the cyclic property of the trace operation one can see that $W$ is a symmetric matrix i.e., $ W_{ji} = W_{ij} $.
Structure of this matrix for the X state \eqref{eq: X-shaped} is
\begin{eqnarray}
W &=& 
\begin{pmatrix}
W_{11} & W_{12} & 0 \\
W_{12} & W_{22} & 0 \\
0 & 0 & W_{33}
\end{pmatrix}
,
\end{eqnarray}
and thus
\begin{eqnarray} \label{eq: LQU-Xstate}
\mathcal{U} &=& 1 - \max \bigg\{ \frac{ W_{11}+W_{22} \pm \sqrt{ [W_{11}-W_{22}]^2 + 4W_{12}^2}  }{2}, W_{33} 
 \bigg\} .
\end{eqnarray}
%


MIN as a measure of nonlocality, a unique property of quantum mechanics, is a description of the maximum global effect caused by locally invariant
measurements \cite{LuFu-PRL-2011}. Originally, MIN was defined in terms of the Hilbert-Schmidt norm $ || X ||_2 = \sqrt{ \tr ( X^{\da}  X) } $ but, then, it was argued that this definition suffers the ancilla and noncontractivity problems. Thus, it was defined based on the trace norm \cite{HuFa-NJP-2015}, Wigner-Yanase skew information \cite{Lietal-EPL-2016} and fidelity \cite{MuSa-PLA-2017}. Trace-norm based MIN avoids problems was mentioned above and is defined via   
\begin{eqnarray} \label{eq: MIN1}
N_1(\rho) &=& \max_{\Pi^{(A)}} || \rho - \Pi^{(A)}(\rho)||_1, \qquad  ||X||_1 = \tr \sqrt{X^{
\da}X} .
\end{eqnarray}
Here, the maximum is taken over the complete set of local projective measurements $ \Pi^{(A)} = \{ \Pi_i^{(A)} \} $, which do not disturb the reduced state $ \rho^{(A)} = \tr_B(\rho) $ i.e., $ \sum_i \Pi_i^{(A)} \rho^{(A)} \Pi_i^{(A)} = \rho^{(A)} $.
Trace norm is invariant under local unitary transformation \cite{CiTuGi-NJP-2014}. By using proper local unitary transformations, one transforms the X state \eqref{eq: X-shaped} into 
\begin{eqnarray} \label{eq: X-shaped-transform}
\rho(t) &=& 
\begin{pmatrix}
\rho_{11} & 0 & 0 & | \rho_{14} | \\
0 & \rho_{22} & | \rho_{23} | & 0 \\
0 & | \rho_{32} | & \rho_{33} & 0 \\
| \rho_{41} | & 0 & 0 & \rho_{44} 
\end{pmatrix}
\end{eqnarray}
which is still X-shaped but now with real elements \cite{EsKhMaDa-OQE-2022}. 
For such type of X-shaped states, it has been shown that the trace-norm MIN \eqref{eq: MIN1} recasts \cite{QiLi-QIP-2016}
\begin{eqnarray} \label{eq: MIN1-Xstate}
N_1 &=&
\begin{cases}
2 ( |\rho_{14}| + |\rho_{23}| ) & x \neq 0 ,  \\
\\
\max \{ |u_1|, |u_2|, |u_3| \}  & x = 0 ,
\end{cases}
\end{eqnarray}
where
\begin{numcases}~
x = \rho_{11} + \rho_{22} - (\rho_{33} + \rho_{44}) , \\
u_1 = 2( | \rho_{14} | + | \rho_{23} | ) , \\
u_2 = 2( - | \rho_{14} | + | \rho_{23} | ) , \\
u_3 = \rho_{11} - \rho_{22} - \rho_{33} + \rho_{44} .
\end{numcases}
%


Another measure of quantum correlations is correlated coherence \cite{Taetal-PRA-2016, Huetal-PR-2018} which is defied as 
\begin{eqnarray} \label{eq: C_cc}
C_{cc} &=& C(\rho) - C( \rho^{(A)} ) - C( \rho^{(B)} ),
\end{eqnarray}
where $ \rho^{(A)} = \tr_B(\rho) $ and $ \rho^{(B)} = \tr_A(\rho) $ are reduced states describing respectively subsystems $A$ and $B$; and $ C( \rho ) $ is the ``quantum coherence" \cite{BaPl-PRL-2014} of the state $ \rho $ quantified by e.g., $\ell_1-$norm which is the sum of the magnitudes of all off-diagonal elements.
The key idea in this definition is to split the total coherence contained in the state $ \rho $ into the local ones, stored locally in the subsystems, and the nonlocal one stored only in the correlated states.
Connection of this notion to quantum discord and entanglement has been argued \cite{Taetal-PRA-2016}. 
For the X state \eqref{eq: X-shaped}, the reduced states are diagonal,
\begin{eqnarray} 
\rho^{(A)} &=& 
\begin{pmatrix}
\rho_{11} + \rho_{22} & 0 \\
0 & \rho_{33} + \rho_{44}
\end{pmatrix}
\label{eq: rhoA-Xstate}
\\
\rho^{(B)} &=&
\begin{pmatrix}
\rho_{11} + \rho_{33} & 0 \\
0 & \rho_{22} + \rho_{44}
\label{eq: rhoB-Xstate}
\end{pmatrix}
\end{eqnarray}
having no coherence. Therefore, for a X-shaped state CC is just the quantum coherence of the global state $ \rho $;
\begin{eqnarray} \label{eq: C_cc-Xstate}
C_{cc} &=& \sum_i \sum_{j \neq i} |\rho_{ij}| = 2 \bigg( |\rho_{14}| + |\rho_{23}| \bigg) .
\end{eqnarray}
Comparison of \eqref{eq: MIN1-Xstate} with \eqref{eq: C_cc-Xstate} reveals that the trace-norm MIN for $x \neq 0 $ is just the CC of the state.

\section{The model, the master equation and its solution} \label{sec: model}

Our model system is a Heisenberg chain with only two localised qubits having the spin-spin interaction 
$ ( J_x S_x^{(1)} S_x^{(2)} + J_y S_y^{(1)} S_y^{(2)} + J_z S_z^{(1)} S_z^{(2)} ) / \hb $ and are in interaction with an applied magnetic field along $z$-direction. 
The Hamiltonian for the $ XY (J_z=0) $ model then recasts,
%
%
%
\begin{eqnarray} \label{eq: ham-XY}
H &=& \frac{J}{\hb} ( S_+^{(1)} \otimes S_-^{(2)} + S_-^{(1)} \otimes S_+^{(2)} ) + \frac{\Del}{\hb} ( S_+^{(1)} \otimes S_+^{(2)} + S_-^{(1)} \otimes S_-^{(2)} ) \nonumber \\
&+& \omega (S_z^{(1)} \otimes \mathds{1}^{(2)} + \mathds{1}^{(1)} \otimes S_z^{(2)} ) ,
\end{eqnarray}
where $ S_{\pm}^{(i)} = S_x^{(i)} \pm i S_y^{(i)} $ are the spin raising (lowering) operators for the $ i^{th} $ qubit and $\om$ is proportional to the intensity of the applied magnetic field. Coupling between qubits have been denoted as $ J = (J_x+J_y)/4 $ and $ \Delta = (J_x-J_y)/4 $ where the latter quantifies the anisotropicity of the interaction.
%
It is noteworthy that systems with two coupled qubits have been utilized extensively in various contexts such as cavity QED \cite{GyMeAw-PRB-2006} and nuclear spin systems with the hope of possible applications in quantum information schemes and quantum computing \cite{NiCh-book-2000, MiLaSaKn-PRA-2002}.

At zero temperature, $T=0$, the state of the system evolves under the Lindbladian master equation \cite{Waetal-JPB-2006}
\begin{eqnarray} \label{eq: master-zero-T}
\dot{\rho}(t) &=& \frac{-i}{\hb} [H, \rho(t)] + \frac{\ga}{\hb^2} ~ \mathcal{D}[S_-^{(1)} \otimes \mathds{1}^{(2)}]\rho(t)
+ \frac{\ga}{\hb^2} ~ \mathcal{D}[\mathds{1}^{(1)} \otimes S_-^{(2)}]\rho(t) ,
%
\end{eqnarray}
where $ \mathcal{D}[A]B = A B A^{\da} - \{A^{\da}A, B \} / 2 $ is the Lindblad super-operator and  $ \{X, Y \} $ denotes the anti-commutator of operators $X$ and $Y$.
In our open quantum system this Lindblad equation describes the population relaxation of the upper state of each qubit with the relaxation rate $\ga$ which for simplicity it has been taken equal for both qubits. In other words, interaction of qubits with the environment is assumed to be the same. 
This assumption is unjustifiable unless the interactions between spins are significantly weaker in comparison to the interaction of spins with the applied magnetic field. In other words, to be a valid assumption, the inter-particle interactions should not significantly change separations between the energy levels.
The single-particle energy levels are $\pm \hb \om/2$ while those of the Hamiltonian \eqref{eq: ham-XY} are given by $\pm \hb J$ and $\pm \hb \Om$ where $\Om=\sqrt{\Del^2 + \om^2}$. Thus, the above requirements imply $ J / \om \ll 1 $ and $ \Del / \om \ll 1 $ or equivalently $ (\Om - \om) / \om \ll 1 $. 
In our calculations, these ratios are taken of the order of $ 10^{-1} $, indicating that the relative variations in the separation between the energy levels are less than 10 \%.

%
The Lindblad equation \eqref{eq: master-zero-T} maintains the shape X of the state i.e., if the initial state is a X-shaped i.e., of the form \eqref{eq: X-shaped} then the evolved state $ \rho(t) $ is too.

%
%
%
%
%
%

In the following, we first examine the evolution of two special X-shaped states, a mixture and the Werner state. Then, we present the steady state solution of the Lindblad equation \eqref{eq: master-zero-T} for the most general X-shaped state \eqref{eq: X-shaped}. It is seen the evolved Werner state and the general steady state are both independent of the interaction intensity $J$. Furthermore, the steady solution is independent of the initial state.
%


Consider the initial state to be an equal mixture of states $ |01\ra $ and $ \frac{1}{\sqrt{2}} (|00\ra+|11\ra) $ i.e.,
\begin{eqnarray} \label{eq: mixture}
\rho_{\text{mixture}} &=& \frac{1}{2} |01 \ra \la 01 | + \frac{1}{4} ( |00 \ra + | 11 \ra )( \la 00 | + \la 11 | ) = 
\nonumber \\
&=&
\begin{pmatrix}
\frac{1}{4} & 0 & 0 & \frac{1}{4} \\
0 & \frac{1}{2} & 0 & 0 \\
0 & 0 & 0 & 0 \\
\frac{1}{4} & 0 & 0 & \frac{1}{4} 
\end{pmatrix}
\end{eqnarray}
Then, the solution of the Lindblad equation \eqref{eq: master-zero-T} is given by
\begin{eqnarray} 
\rho_{11}(t) &=& \frac{ 1 }{ 4 \Om   (\ga^2 + 4 \Om^2 ) } \bigg\{ 4 \Om \Del^2 +
e^{-2 \ga  t} \Om  \bigg[ \ga^2 + 4 \om  (\Del +\om ) \bigg ] 
\nonumber \\
&~& + 2 \Del  e^{- \ga  t} \bigg[ \ga  (\om -2 \Del ) \sin (2 \Om t )-2 \om  \Om  \cos (2 \Om t) \bigg] \bigg\}
\label{eq: rhot_mix_11}
\\ 
\rho_{14}(t) &=& \frac{ 1 }{ 4 \Om^3 (\ga^2 + 4 \Om^2 ) } \bigg\{ - 8 \Del \om \Om^3 
+ e^{-\ga t} \om  \Om  \bigg( \ga^2 ( \om - 2 \Del ) + 4 \om  \Om^2 \bigg) \cos (2 \Om t)
\nonumber \\
&~& + \Del \Om  e^{-\ga t} \bigg( (\ga^2 + 4 \Om^2) ( \Del + 2 \om ) 
+ 4 \ga  \om  \Om \sin (2 \Om t) \bigg) 
\bigg\}
\nonumber \\
&& - i \frac{ 4 \ga \Del \Om + e^{-\ga  t}  \bigg[  \bigg( \ga ^2 ( \om -2 \Del ) + 4 \om  \Om^2 \bigg) \sin (2 \Om t ) -4 \ga  \Del  \Om  \cos (2 \Om t) \bigg] }{ 4 \Om (\ga ^2+4 \Om ^2 ) }
 \label{eq: rhot_mix_14}
\\
\rho_{22}(t) &=& \frac{ 1 }{ 4 \Om^3 (\ga^2 + 4 \Om^2 ) } \bigg\{
4 \Del^2 \Om^3 
+ e^{-\ga  t} \bigg[ \Om ^3 ( \ga^2 + 4 \Om ^2 ) \cos (2 J t)
\nonumber \\
&~& 
+\ga  \Del  \bigg( \ga  \Om  (2 \Del -\om ) \cos (2 \Om t)-2 \om  \Om ^2 \sin (2 \Om t ) \bigg) \bigg]  
\nonumber \\
&~& 
+ e^{-\ga  t} \Om  \bigg[ -\Om ^2 ( \ga ^2+4 \om  (\Del +\om ) ) + \om  ( \ga^2+4 \Om ^2 ) ( \Del +2 \om ) e^{\ga  t} \bigg] 
\bigg\}
\label{eq: rhot_mix_22}
\\
\rho_{23}(t) &=& \rho_{32}^*(t) = \frac{i}{4} e^{-\ga t} ~ \sin(2 J t)
\label{eq: rhot_mix_23}
\\
\rho_{33}(t)  &=& \rho_{22}(t) - \frac{1}{2} e^{-\ga t} \cos (2 J t)
\\
\rho_{41}(t) &=& \rho_{14}^*(t) 
\label{eq: rhot_mix_41}
\\
\rho_{44}(t) &=& 1 - [ \rho_{11}(t) + \rho_{22}(t) + \rho_{33}(t) ]
\label{eq: rhot_mix_44}
\end{eqnarray}
where we have defined
\begin{eqnarray} \label{eq: Om}
\Om &=& \sqrt{ \Delta^2 + \om^2 } .
\end{eqnarray}
One notes that only elements $ \rho_{22}(t) $, $ \rho_{23}(t) $,$ \rho_{32}(t) $ and $ \rho_{33}(t) $ depends on the parameter $J$ which due to the damping factor $ e^{-\ga t} $, this dependence dies at the limit $ t \to \infty $.
%


Take the initial state to be now the Werner state,
\begin{eqnarray} \label{eq: Werner}
\rho_W &=& 
p | \Psi_- \ra \la \Psi_- | + \frac{1-p}{4} \mathds{1} \otimes \mathds{1} =
\begin{pmatrix}
\frac{1-p}{4} & 0 & 0 & 0 \\
0 & \frac{1+p}{4} & -\frac{p}{2} & 0 \\
0 & -\frac{p}{2} & \frac{1+p}{4} & 0 \\
0 & 0 & 0 & \frac{1-p}{4} 
\end{pmatrix}
\end{eqnarray}
where $ | \Psi_- \ra = \frac{1}{\sqrt{2}}( | 01 \ra - | 10 \ra) $ is one of the maximally entangled Bell states, and $p$ a real number with constraint $ p \in [-\frac{1}{3}, 1 ] $ which comes from the non-negativity requirement of the density matrix.  
For $ p \in [-\frac{1}{3}, \frac{1}{3} ] $ the Werner state is separable and for $ p \in [-\frac{1}{3}, \frac{1}{ \sqrt{2} } ] $ it is local \cite{Luc-thesis-2019}. From this one sees that the Werner state is entangled but local for $ p \in (\frac{1}{3}, \frac{1}{ \sqrt{2} } ] $ i.e., entanglement does not imply nonlocality.
The purity of the Werner state \eqref{eq: Werner} is $ \tr(\rho_W^2) = (1+3p^2)/4 $ being one for $ p = 1 $ and $1/4$ for $p=0$. The former implies the pure ensemble $ | \Psi_- \ra \la \Psi_- | $ while the latter corresponds to the completely random ensemble.  
The solution of the Lindblad equation \eqref{eq: master-zero-T} this time is given by
\begin{eqnarray} 
\rho_{11}(t) &=& \frac{ 4 \Del^2 \Om - 4\Del^2 \ga e^{-\ga t} \sin(2\Om t) + 
\Om e^{-2\ga t} (4\om^2 - 4\Om^2 p + \ga^2(1-p)) }{ 4\Om(\ga^2+4\Om^2) }
\label{eq: rhot_W_11}
\\ 
\rho_{14}(t) &=&  \frac{i \Del }{2 \Om ^3 (\ga^2 + 4 \Om^2 )} 
\bigg \{ - 2 (\ga -2 i \om )\Om^3 + e^{-\ga t} \bigg[ - 4 i \om  \Om^3 -2 i \ga^2 \om  \Om \sin^2(\Om t)
\nonumber \\
&~& 
  + \ga^2 \Om^2 \sin(2 \Om t ) + 2 \ga  \Om^2 \bigg(\Om  \cos (2 \Om t) - i \om  \sin (2 \Om t)\bigg) \bigg] \bigg\} \label{eq: rhot_W_14}
\\
\rho_{22}(t) &=& \frac{1}{4 \Om ^2 (\ga ^2+4 \Om ^2 )} 
\bigg \{ 4 \Om^2 \bigg[ \Del ^2+e^{-2 \ga  t} (p \Del ^2 + (p-1) \om ^2 )+2 \om ^2 e^{-\ga t} \bigg] 
\nonumber \\
&~& + \ga^2 \bigg[ (p-1) \Om ^2 e^{-2 \ga  t}+2 e^{-\ga t} (\Del ^2 \cos (2 \Om t)+\om^2 ) \bigg]
\bigg\}
\label{eq: rhot_W_22}
\\
\rho_{23}(t) &=& \rho_{32}(t) = -\frac{p}{2} e^{-\ga t} 
\label{eq: rhot_W_23}
\\
\rho_{33}(t)  &=& \rho_{22}(t)
\\
\rho_{41}(t) &=& \rho_{14}^*(t)
\\
\rho_{44}(t) &=& \frac{1}{ 4 \Om^3 (\ga^2 + 4 \Om^2 ) } \bigg\{ 4 \Om^3 ( \ga^2 + 3 \om^2 + \Om^2 ) 
\nonumber\\
&~&
- \Om^3 e^{-2 \ga  t} \bigg[ (\ga^2 + 4 \Om^2) p  - (\ga^2 + 4 \om^2) \bigg] - 4 \om^2 \Om (\ga^2 + 4 \Om^2 ) e^{-\ga t} 
\nonumber\\
&~& 
 + 4 \ga \Del^2 e^{-\ga t} \bigg[ \Om^2 \sin (2 \Om t ) - \ga \Om  \cos(2 \Om t) \bigg] \bigg \} 
\label{eq: rhot_W_44}
\end{eqnarray}
where $\Om$ is given by \eqref{eq: Om}.
As one sees the state $ \rho_W(t) $ is independent of the parameter $J$ appearing in the Hamiltonian being responsible for the isotropic part of the spin-spin interaction.
An interesting point about the Werner state is that for $ \ga = 0 $, the density operator is independent of time, which is the manifestation of the fact that the Werner state \eqref{eq: Werner} commutes with the Hamiltonian \eqref{eq: ham-XY} of the system, $ [H, \rho_W(t)] = 0 $.


Although the master equation \eqref{eq: master-zero-T} can be analytically solved for the general X-shaped state \eqref{eq: X-shaped} but since the relations are lengthy we only present the steady state solution i.e., the state at $ t \to \infty $ which is
\begin{eqnarray} \label{eq: rho-steady}
\rho_{\infty} &=& 
\begin{pmatrix}
\frac{ \Del^2 }{\ga^2 + 4 \Om^2} & 0 & 0 & - \frac{\Del(2\om + i \ga)}{\ga^2 + 4 \Om^2} \\
0 & \frac{ \Del^2 }{\ga^2 + 4 \Om^2} & 0 & 0 \\
0 & 0 & \frac{ \Del^2 }{\ga^2 + 4 \Om^2} & 0 \\
- \frac{\Del(2\om - i \ga)}{\ga^2 + 4 \Om^2} & 0 & 0 & \frac{ \ga^2 + 3\om^2 + \Om^2 }{\ga^2 + 4 \Om^2}
\end{pmatrix}
.
\end{eqnarray}
%
This result show that the steady state is the same for any X-shaped state and thus all such states have the same steady state quantum correlations.
Note that the condition $ \rho_{22} \rho_{33}  < | \rho_{14} |^2 $ then yields to
\begin{eqnarray} \label{eq: Entangle-con-steady}
| \Del | < \sqrt{\ga^2 + 4 \om^2}
\end{eqnarray}
for the steady state to be entangled.
As we saw, reduced density operators for a X-shaped state are both always diagonal with zero quantum coherence. However, Eq. \eqref{eq: rho-steady} shows that the global steady state $\ell_1$-norm coherence is, 
\begin{eqnarray}
C_{\ell_1}(\rho_{\infty}) &=& \sum_i \sum_{j \neq i} | \rho_{\infty, ij} | = \frac{ | \Del | \sqrt{ 4\om^2 + \ga^2 } }{\ga^2 + 4 \Om^2} ,
\end{eqnarray}
i.e., decoherence does not eliminate the off-diagonal elements of the global state.

\section{Temperature dependence of the steady state solution} \label{sec: temp}

Now we aim to consider the behaviour of steady state quantum correlations with temperature. To this end, we assume both qubits interact with the same thermal bath and consider the master equation
\begin{eqnarray} \label{eq: master-temp}
\dot{\rho}(t) &=& \frac{-i}{\hb} [H, \rho(t)] + \frac{\ga}{\hb^2} (\bar{n}+1) ~ \mathcal{D}[S_-^{(1)} \otimes \mathds{1}^{(2)}]\rho(t) + \frac{\ga}{\hb^2} (\bar{n}+1) ~ \mathcal{D}[\mathds{1}^{(1)} \otimes S_-^{(2)}]\rho(t) 
\nonumber \\
&~& + \frac{\ga}{\hb^2} \bar{n} ~ \mathcal{D}[S_+^{(1)} \otimes \mathds{1}^{(2)}]\rho(t) + \frac{\ga}{\hb^2} \bar{n} ~ \mathcal{D}[\mathds{1}^{(1)} \otimes S_+^{(2)}]\rho(t) , 
\end{eqnarray}
$ \bar{n} $ being the average excitation of the thermal bath parametrizing the temperature: being zero for $T=0$ \cite{Waetal-JPB-2006}. This equation is known as quantum optical master equation \cite{BrPe-book-2002}\footnote{page: 148, equation (3.219)}
%
where the first two terms of the non-unitary part in the right side of the equation, describe the decay process while the remaining part i.e., the last two terms on the right hand side, are responsible to the excitation process with a rate that depends on temperature through $\bar{n}$. Note that in the limit of zero temperature, this equation reduces to Eq. \eqref{eq: master-zero-T} which describes the spontaneous decay, a purely dissipative process \cite{MiCaKuBu-PR-2005}.
Eq. \eqref{eq: master-temp} can be analytically solved for the X-state \eqref{eq: X-shaped}. Again, one sees that this master equation preserves the shape of the state. Here, we are only interested in the steady state, $t \to \infty $, solution which is given by
\begin{numcases}~
\rho_{11}(\infty) = \frac{ (2 \bar{n}+1)^2  (\Del^2 + \ga^2 \bar{n}^2 ) + 4 \bar{n}^2 \om^2 } { (2 \bar{n}+1)^2 [ 4 \Om^2 + \ga^2 (2 \bar{n}+1)^2 ] } , \label{eq: rho11_temp}
\\
\rho_{14}(\infty) = -\frac{ 2 \om \Del + i \ga \Del ( 2 \bar{n} + 1 ) }{ (2 \bar{n}+1) [ 4 \Om^2 + \ga^2 (2 \bar{n}+1)^2 ] } , \label{eq: rho14_temp}
\\
\rho_{22}(\infty) = \frac{ \Del^2 + \bar{n} ( \bar{n}+1) (4 \Om^2 + \ga^2 ( 2\bar{n} + 1 )^2 ) } { (2 \bar{n}+1)^2 [ 4 \Om^2 + \ga^2 (2 \bar{n}+1)^2 ] } , \label{eq: rho22_temp}
\\
\rho_{23}(\infty) = \rho_{32}(\infty) = 0 , \label{eq: rho23_temp}
\\
\rho_{33}(\infty) = \rho_{22}(\infty) , \label{eq: rho33_temp}
\\
\rho_{41}(\infty) = \rho_{14}^*(\infty) , \label{eq: rho41_temp}
\\
\rho_{44}(\infty) =\frac{ ( 2 \bar{n} + 1 )^2 (\Del^2 + \ga^2 ( \bar{n} + 1 )^2 ) + 4 ( \bar{n} + 1 )^2 \om^2 } { (2 \bar{n}+1)^2 [ 4 \Om^2 + \ga^2 (2 \bar{n}+1)^2 ] } .
\label{eq: rho44_temp}
\end{numcases}
%
%
%
%
%
%
%
From these equations one obtains 
\begin{eqnarray} 
\mathcal{C}(\infty) &=& 2 \max \bigg\{0,
\frac{ (2 \bar{n}+1)  | \Del | \sqrt{ 4 \om^2 + \ga^2 (2 \bar{n}+1)^2 } - \Del^2} { (2 \bar{n}+1)^2 [ 4 \Om^2 + \ga^2 (2 \bar{n}+1)^2 ] } - \frac{ \bar{n}( \bar{n} + 1 ) }{ (2 \bar{n}+1)^2 } \bigg\} ,
\label{eq: Con-st-temp}
\\
C_{cc}(\infty) &=& 2 \frac{ | \Del | \sqrt{ 4 \om^2 + \ga^2 ( 2\bar{n}+1 )^2 } } { (2 \bar{n}+1) [ 4 \Om^2 + \ga^2 (2 \bar{n}+1)^2 ] } , \label{eq: Ccc-st-temp}
 \\
N_1(\infty) &=& C_{cc}(\infty) ,
\end{eqnarray}
respectively for the steady state concurrence, CC and MIN.
In this case, the $W$ matrix \eqref{eq: W-mat} is diagonal with $ W_{11}(\infty) = W_{22}(\infty) $. Thus, for the steady state LQU one has that 
\begin{eqnarray} \label{eq: LQU-st-temp}
\mathcal{U}(\infty) &=& \max \{ W_{11}(\infty), W_{33}(\infty) \}
\end{eqnarray}
For $ \bar{n} = 0 $ one obtains
\begin{eqnarray}
W_{11}(\infty) &=& \frac{ \sqrt{2} | \Del | }{ \ga^2 + 4 \Om^2 }
\bigg\{
\sqrt{ \ga^2 + 2 \om^2 + 2 \Om^2 - \sqrt{ (\ga^2 + 4 \om^2)(\ga^2 + 4 \Om^2) } }
\nonumber \\
&~& + \sqrt{ \ga^2 + 2 \om^2 + 2 \Om^2 + \sqrt{ (\ga^2 + 4 \om^2)(\ga^2 + 4 \Om^2) } }
\bigg\} ,
 \label{eq: Wmat-st} 
 \\
W_{33}(\infty) &=&  \frac{ \ga^4 + 4 \ga^2 ( \om^2 + \Om^2 ) + 16 ( \Del^4 + \om^2 \Om^2 ) }{ ( \ga^2 + 4 \Om^2 )^2 } .
\end{eqnarray}

For non-interacting qubits, $ J = \Del = 0 $, the steady state density matrix recasts to the diagonal form,
\begin{eqnarray} \label{eq: rhost-temp}
\rho(\infty) &=& 
\text{diag}\left\{ 
 \frac{\bar{n}^2}{(2\bar{n}+1)^2}, 
 \frac{\bar{n}(\bar{n}+1)}{(2\bar{n}+1)^2},
 \frac{\bar{n}(\bar{n}+1)}{(2\bar{n}+1)^2}, 
  \frac{(\bar{n}+1)^2}{(2\bar{n}+1)^2}
\right\}
\end{eqnarray}
implying that all coherences are zero. Then, from \eqref{eq: rho_Dicke} one obtains
\begin{equation} \label{eq: rhost-temp-Dicke}
\varrho_{ee}(\infty) = \frac{\bar{n}^2}{(2\bar{n}+1)^2}, \quad \varrho_{gg}(\infty) = \frac{(\bar{n}+1)^2}{(2\bar{n}+1)^2}, 
\quad \varrho_{ss}(\infty) = \varrho_{aa}(\infty) = \frac{\bar{n}(\bar{n}+1)}{(2\bar{n}+1)^2}
\quad \varrho_{sa}(\infty) = \varrho_{eg}(\infty) = 0
\end{equation}
for the elements of the density matrix in the Dicke basis. These relations reveal that all steady-state coherences vanish; and the symmetric and antisymmetric states are equally populated in the stationary regime.  
Then, from Eqs. \eqref{eq: C1-Dicke}, \eqref{eq: C2-Dicke} and \eqref{eq: concur} it is found that for non-interacting qubits, the steady state concurrence is zero. 


\section{Results and discussion} \label{sec: results}

In this section, we aim to consider the impact of the relaxation rate $\ga$, responsible for the interaction of our qubit-qubit Heisenberg XY system with the environment, on the dynamics of quantum correlations by fixing other parameters of the problem including coupling strengths $J$ and $\Del$; and magnetic field intensity $\om$. However, as we saw in the previous section, Werner states and thus corresponding quantum correlations are independent of $J$. The same is valid for the general X-shaped state in the steady situation. Furthermore, in some calculations we fix all parameters other than $\Del$ to study anisotropicity dependence of quantum correlations. 
Our investigation will primarily focus on examining how the zero-temperature steady state quantum correlations appear for different initial X states, including a mixture and the Werner state. 
After, the interesting phenomenon of entanglement sudden death, for independent qubits, will be considered for two initially entangled and afterwards not interacting qubits both for zero and finite temperatures. The time of ESD will be obtained analytically (numerically) for zero (finite) temperatures.
Environment-induced quantum correlations is another phenomenon which will be considered. 
At the end finite temperature quantum correlations will be examined in the steady state case by studying their behaviour with temperature.  

\begin{figure} 
\centering
\includegraphics[width=12cm,angle=-0]{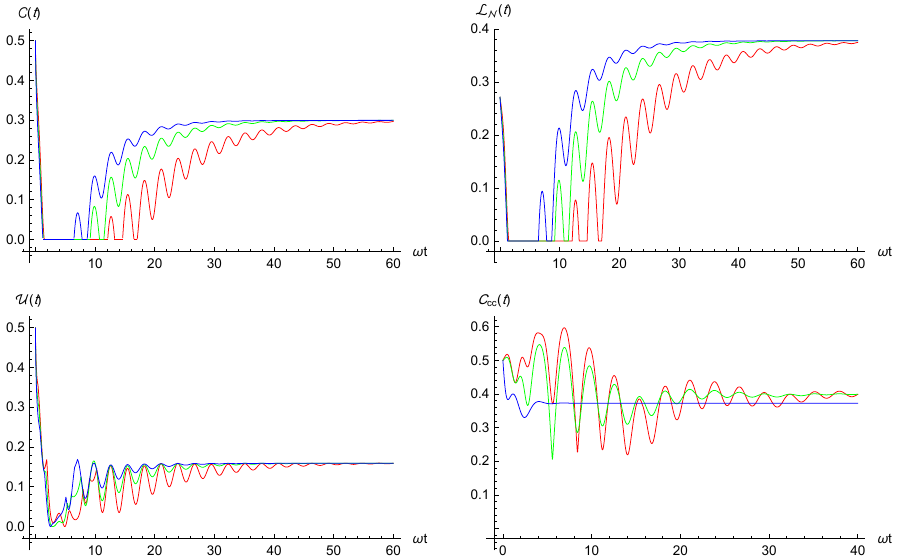}
\caption{
Evolution of the quantum correlations, concurrence (left top panel), log negativity (right top panel), LQU (left bottom panel) and CC (right bottom panel), for $ \ga = 0.1 \om $ (red curves), $ \ga = 0.15 \om $ (green curves), $ \ga = 0.2 \om $ (blue curves) for $ J = 0.1 \om $ and $ \Del = 0.5 \om $. The quantum system is initially described by the mixture \eqref{eq: mixture}. 
}
\label{fig: cor-mixture-t} 
\end{figure}

Figure \ref{fig: cor-mixture-t} depicts evolution of different quantum correlations for $ J = 0.1 \om $ and $ \Del = 0.5 \om $ for three different values of the relaxation rate $\ga = 0.1 \om$, $\ga = 0.15 \om$ and $\ga = 0.2 \om$. Here, system is initially described by the entangled mixture \eqref{eq: mixture}.  Initial values of the quantum correlations, being independent of $\ga$, are $ \mathcal{C}(0) = 0.5 $, $ \mathcal{L_N}(0) = 0.2716 $, $ \mathcal{U}(0) = 0.5 $ and $ C_{cc}(0) = 0.5 $. 
As this figure reveals, both entanglement quantifiers, concurrence and log negativity, exhibit 
several dark and revival periods \cite{FiTa-PRA-2006} before reaching the steady state. 
Because of the spontaneous emission, the initial entanglement diminishes with time, disappears in a given finite time which itself depends on the value of $\ga$ and after a dark while it revives.
The number and length of these dark intervals both depend on the value of $\ga$. The first dark period is the longest one, which itself decreases with $\ga$.
%
%
The difference between entanglement quantifiers, $ \mathcal{C} $ and $ \mathcal{L_N} $, and quantum correlations beyond entanglement, $ \mathcal{U} $ and $C_{cc}$, is that the former ones revive after zero entanglement, while the others can revive immediately several times without completely vanishing for a period of time; there are no dark periods for quantum correlations beyond QE.
Moreover, LQU captures less quantum correlations than the other quantifiers, with the former converging to a stationary value of $ 0.1597 $ for $\ga = 0.1 \om $.
As one expects, the transition to the steady state occurs sooner for higher values of $\ga$ and as one can check analytically from Eq. (\ref{eq: rho-steady}) the steady state value of quantum correlation although is independent of the initial state but depends on the value of $ \ga $. 
%

%

\begin{figure} 
\centering
\includegraphics[width=12cm,angle=-0]{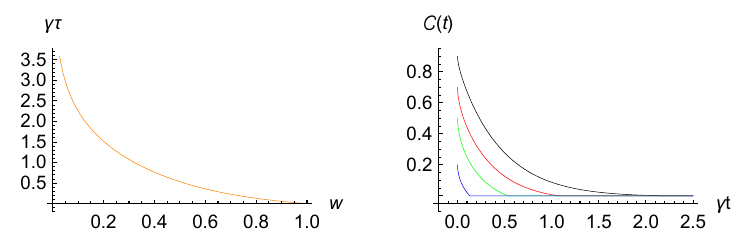}
\caption{
Left panel: time of entanglement death $ \ga \tau $ given by Eq. \eqref{eq: ESD-time} in terms of the weight $w$ appearing in the mixture \eqref{eq: mix-nonequal-0}.
Right panel: evolution of concurrence \eqref{eq: concur-mix-nonequal} for $ w = 0.1 $ (black curve), $ w = 0.3 $ (red curve), $ w = 0.5 $ (green curve) and $ w = 0.8 $ (blue curve).  
Parameters have been chosen $ \Del =  J = 0 $ i.e, qubits solely interact with the common environment, not with each other. 
}
\label{fig: pDeath} 
\end{figure}

We consider now a system of two independent non-interacting qubits, $J = \Del = 0$, which initially is described by the mixture 
%
\begin{eqnarray} \label{eq: mix-nonequal-0}
\rho_{\text{mix}} &=& w |01 \ra \la 01 | + \frac{1-w}{2} ( |00 \ra + | 11 \ra )( \la 00 | + \la 11 | ), \qquad 0 \leq w \leq 1. 
\end{eqnarray}
Then one has that $ \xi = \tr( \rho_{\text{mix}}^2 ) = 1 - 2w(1-w) $ for the purity of the state.
Time dependent density matrix obtained from the solution of the Lindblad equation \eqref{eq: master-zero-T} with $J = \Del = 0$ and the initial condition \eqref{eq: mix-nonequal-0} is then,
\begin{numcases}~ 
\rho_{11}(t) = \frac{1-w}{2} e^{-2\ga t} \label{eq: mix-non-t-11} \\
\rho_{14}(t) = \frac{1-w}{2} e^{- ( \ga + 2 i \om )t} \label{eq: mix-non-t-14} \\
\rho_{22}(t) = e^{-3\ga t/2} \bigg( \sinh( \frac{\ga t}{2} ) + w \cosh( \frac{\ga t}{2} ) \bigg) \label{eq: mix-non-t-22} \\
\rho_{23}(t) = \rho_{32}(t) = 0 \label{eq: mix-non-t-23} \\
\rho_{33}(t) = (1-w) e^{-3\ga t/2} \sinh( \frac{\ga t}{2} ) \label{eq: mix-non-t-33} \\
\rho_{41}(t) = \frac{1-w}{2} e^{- ( \ga - 2 i \om )t} \label{eq: mix-non-t-41} \\
\rho_{44}(t) = \frac{1-w}{2} e^{-2\ga t} + 2 e^{-\ga t/2} \sinh( \frac{\ga t}{2} ) \label{eq: mix-non-t-44}
\end{numcases}
It follows that in the stationary regime, $ t \to \infty $, all coherences vanish and all populations except $\rho_{44} $ become zero. This implies that, in this limit, the non-interacting qubits accumulate in the ground state as a result of spontaneous decay due to interaction with the environment. Note that, as we have shown in Eq. \eqref{eq: rho-steady}, this behaviour is independent of the initial state.
Now, we consider evolution of concurrence. For this state, $ \mathcal{C}_2(t) $ given by \eqref{eq: C2} is negative, having no contribution to the entanglement; only the two-photon coherence $ \mathcal{C}_1(t) $ is present. 
Thus, from the definition \eqref{eq: concur} one obtains
\begin{equation} \label{eq: concur-mix-nonequal}
\mathcal{C}(t) = \max \bigg\{0, e^{-\ga t} \bigg[ 1 - w - 2 e^{-\ga t/2} \sqrt{ (1-w) \sinh( \frac{\ga t}{2} ) \bigg( \sinh( \frac{\ga t}{2} ) + w \cosh( \frac{\ga t}{2} ) \bigg)  } \bigg]  \bigg\}
\end{equation}
for the concurrence of this mixture. This equation shows that this state is always separable for $w=1$ as it is obvious from \eqref{eq: mix-nonequal-0} for the initial state. 
For $ 0 \leq w < 1 $ which the initial state is entangled, Eq. (\ref{eq: concur-mix-nonequal}) shows that there is a finite time when the entanglement disappears, the signature of ESD. 
This time is obtained as
\begin{eqnarray} \label{eq: ESD-time}
\tau &=& \frac{1}{\ga} \ln \bigg( \frac{1 + \sqrt{1-2w(1-w)}}{2w} \bigg)
\end{eqnarray}
for the time of ESD which is infinite for $ w = 0 $ meaning that the entanglement does not die suddenly for the maximally entangled state $ \frac{1}{2} ( |00 \ra + | 11 \ra )( \la 00 | + \la 11 | ) $. The age of entanglement is shorter for higher values of both $\ga$ and $w$. See figure \ref{fig: pDeath}. The weight factor $w$ is itself related to purity of the state \eqref{eq: mix-nonequal-0} through $ \xi = \tr(\rho_{\text{mix}}^2) = 1 - 2 w (1-w) $. As a result, by engineering the purity of the initial state, ESD can be slowed down.

The origin of ESD can be understood with regard to the populations and coherences; and the rates with which these quantities decay.
We have considered non-interacting qubits for which $ J = 0 = \Del $. Thus, the Hamiltonian \eqref{eq: ham-XY} recasts in $ H = \om ( |00\ra\la 00| - |11\ra\la 11| ) $ with eigenvalues $ \{\om, 0, 0, -\om \}$ and corresponding eigenvectors $ | 00 \ra $, $ | 01 \ra $, $ | 10 \ra $ and $ | 11 \ra $. From the linear superpositions of $ | 0 1 \ra $ and $ | 1 0 \ra $, one then constructs symmetric and antisymmetric eigenvectors. In this way, four vectors of the Dicke basis \eqref{eq: ee}, \eqref{eq: gg}, \eqref{eq: symmetric} and \eqref{eq: anti-symmetric} are obtained.
ESD is then explained by analysing Eq. \eqref{eq: C1-Dicke} in terms of the populations of the symmetric and antisymmetric states, $\varrho_{ss}(t)$ and $\varrho_{aa}(t)$, and the rates with which the populations and the two-photon coherence $|\varrho_{eg}(t)|$ decay \cite{FiTa-JCMSE-2010}.

\begin{figure} 
\centering
\includegraphics[width=12cm,angle=-0]{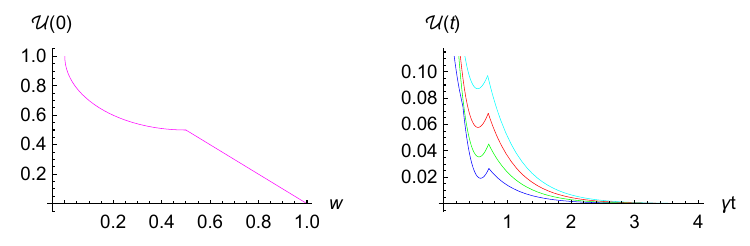}
\caption{
Initial value of LQU in terms of the probability weight $w$ (left panel) and evolution of LQU (right panel) for $ w = 0.4 $ (cyan), $ w = 0.5 $ (red), $ w = 0.6 $ (green) and $ w = 0.7 $ (blue) for $ \Del =  J = 0 $ i.e, qubits solely interact with the common environment not with each other. The system is initially described by the mixture \eqref{eq: mix-nonequal-0}.
}
\label{fig: pLQUED1} 
\end{figure}

%
Now the natural question arises: Does the phenomenon of death occurs for correlations beyond entanglement? From, Eq. \eqref{eq: C_cc-Xstate} we have
\begin{eqnarray} \label{eq: Ccc-ED}
C_{cc}(t) &=& (1-w) e^{-\ga t}
\end{eqnarray}
for CC or equivalently for MIN with respect to \eqref{eq: MIN1-Xstate}. 
This equation shows that $ C_{cc}(0) = 1 - w $, being nonzero for $ w \neq 1 $, and CC decreases exponentially; the usual behaviour of decaying systems. Thus, there is no phenomenon of CC sudden death, i.e., this quantity does not disappear in a finite time.
For LQU which is computed by \eqref{eq: LQU-Xstate} analytical calculations show that the square root expression appearing in this equation is zero. Thus, LQU for $ J = \Del = 0 $ is obtained from
\begin{eqnarray} \label{eq: LQU-ED}
\mathcal{U}(t) &=& \max \bigg\{ \frac{ W_{11}(t)+W_{22}(t)}{2}, W_{33}(t) \bigg\}
\end{eqnarray}
where $W_{ij}$ is given by \eqref{eq: W-mat}. The initial value of LQU is $ \mathcal{U}(0) = \max \{1-w, 1-\sqrt{w(1-w)} \} $ being zero for $ w=1 $. 
In the left panel of Fig. \ref{fig: pLQUED1} we have plotted initial LQU versus the probability weight. 
As the numerical computations in the right panel of Fig. \ref{fig: pLQUED1} show, LQU decreases first reaching a local minimum then increases to a given value and breaks downwards and continuously, not abruptly, vanishes. Ref. \cite{GuPeZeWa-QIP-2020} has already reported this sudden change in LQU when one of the qubits is subjected to local decoherence noise. In any way, there is no LQU sudden death.
%
%
Note that breaking points in the curves are due to the ``max" operation in the definition \eqref{eq: LQU-ED}.

\begin{figure} 
\centering
\includegraphics[width=12cm,angle=-0]{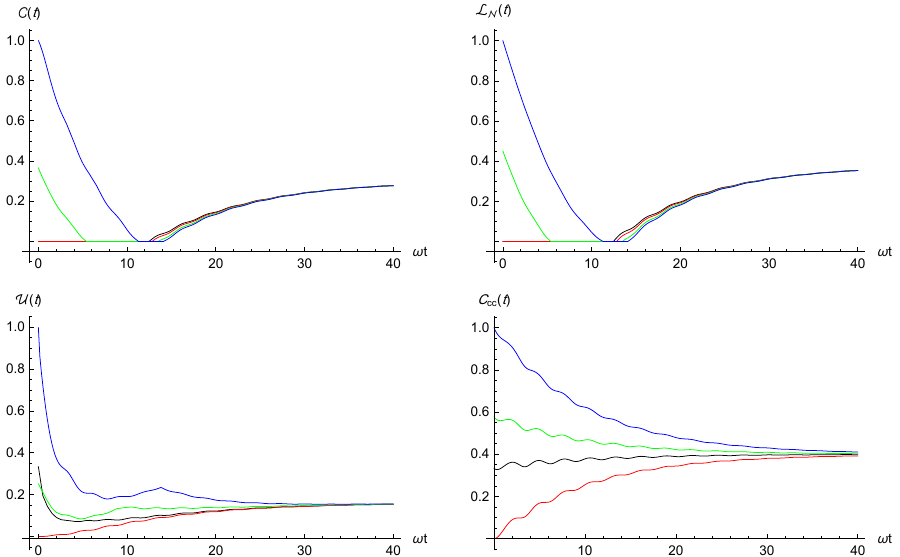}
\caption{
Evolution of the quantum correlations, concurrence (left top panel), log negativity (right top panel), LQU (left bottom panel) and CC (right bottom panel), for $ p = -\frac{1}{3} $ (black curves), $ p = 0 $ (red curves), $ p = \frac{1}{ \sqrt{3} } $ (green curves), and $ p = 1 $ (blue curves) for  $ \ga = 0.1 \om $ and $ \Del = 0.5 \om $.  Note that the Werner state is independent of the isotropic part of qubits' interaction, i.e., the coefficient $J$. 
}
\label{fig: cor-Werner-t} 
\end{figure}

Next, we illustrate in figure \ref{fig: cor-Werner-t} the dynamics of quantum correlations for the Werner state \eqref{eq: Werner} for different values of $ p $. 
Initial values of quantum correlations are $ \mathcal{C}(0) = \max\{ 0, |p| - \frac{1-p}{2} \} $, $ \mathcal{L_N}(0) = \log_2 \left( \frac{1+p}{2} + \frac{1}{2} \left| \frac{1-p}{2} - |p| \right| + \frac{1}{2} \left| \frac{1-p}{2} + |p| \right| \right)  $, $ C_{cc}(0) = |p| $ and $ \mathcal{U}(0) = 1 - \frac{1}{2} [ 1 - p + \sqrt{(1-p)(1+3p)} ] $ from which one finds the concurrence is non-zero only for $ \frac{1}{3} < p \leq 1 $ implying, as is well-known, that in this region the Werner state \eqref{eq: Werner} is entangled.
While QE, quantified by both concurrence and log negativity, is initially zero for values $ p = -\frac{1}{3} $ and $ p = 0 $ after a while it is born and finally reaches the steady value $ 0.2999 $  and $ 0.3784 $, being independent of $p$, respectively for concurrence and log negativity for $ \ga = 0.1 \om $ and $ \Del = 0.5 \om $.
Using Eq. (\ref{eq: rho-steady}) one obtains $ \mathcal{C}(\infty) = 2 \max \left\{0, \frac{ | \Del | \sqrt{ 4 \om^2 + \ga^2 } - \Del^2} { 4 \Om^2 + \ga^2 } \right\} $ for the steady state value of quantum concurrence which is non-zero only for $ |\Del| < \sqrt{ 4 \om^2 + \ga^2 } $ . As pointed out in the previous section, for $ \ga = 0 $, the Werner state \eqref{eq: Werner} does not evolve in time, meaning that all quantum correlations remain constant. From all these we see that if both conditions $ \frac{1}{3} < p \leq 1 $ and  $ |\Del| < \sqrt{ 4 \om^2 + \ga^2 } $ fulfil simultaneously then for $ |p| - \frac{1-p}{2} < \frac{ | \Del | \sqrt{ 4 \om^2 + \ga^2 } - \Del^2} { 4 \Om^2 + \ga^2 } $, the steady state is more entangled than the initial state. For parameters $ \ga = 0.1 \om $ and $ \Del = 0.5 \om $ where $ \mathcal{C}(\infty) = 0.2999 $, this condition fulfils for $ \frac{1}{3} < p < 0.5333 $. 
%
%
In contrast to the entanglement which is generated after a while, as can be found from the red curve corresponding to $p=0$, both LQU and CC are generated much more quickly than entanglement. As before, steady state LQU captures less quantum correlations than the other quantifiers. Note that the Werner state for $p=0$ is the completely unpolarized state.

\begin{figure} 
\centering
\includegraphics[width=12cm,angle=-0]{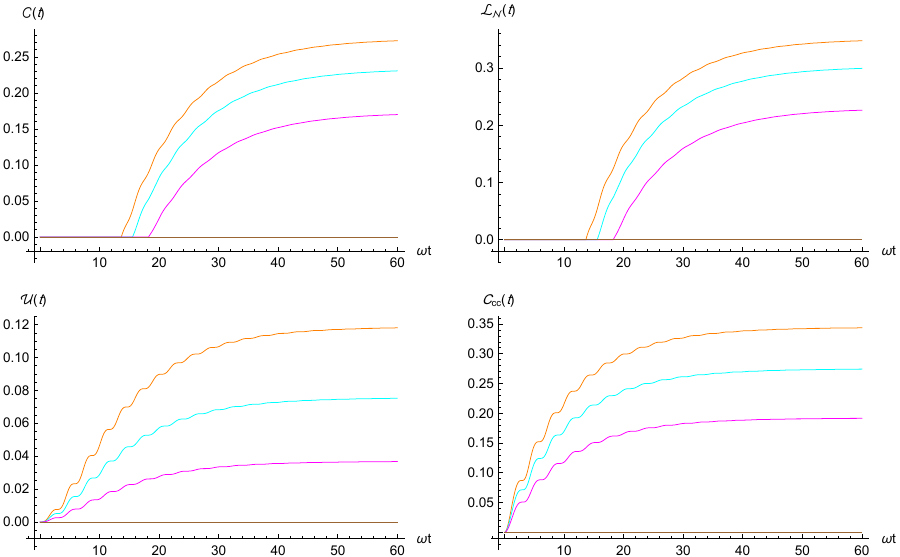}
\caption{
Evolution of the quantum correlations, concurrence (left top panel), log negativity (right top panel), LQU (left bottom panel) and CC (right bottom panel) for different values of the anisotropy interaction coefficient: $ \Del = 0 $ (brown), $ \Del = 0.2 \om $ (magenta) $ \Del = 0.3 \om $ (cyan) and $ \Del = 0.4 \om $ (orange); and for $ \ga = 0.1 \om $. System is initially described by the Werner state \eqref{eq: Werner} with $ p = 0 $ (completely random ensemble; $\rho_0 = \frac{1}{4} \mathds{1}_4 $).
}
\label{fig: cor-Werner-t-separ}
\end{figure}

After, we highlight the constructive role played by the environment in favour of quantum correlations; the so-called environment-induced or decoherence-induced quantum correlations. To this end, we consider our system is initially described by the Werner state \eqref{eq: Werner} with $ p = 0 $ i.e., $\rho_0 = \frac{1}{4} \mathds{1}_4 $; the completely random ensemble which contains no correlations. 
Then, evolution of quantum correlations is considered for different values of the interaction strength $\Del$. Note that the evolved state and thus quantum correlations are all independent of the value of $J$. 
Figure \ref{fig: cor-Werner-t-separ} depicts variation of quantum correlations with time for a given value of $\ga$ but different values of interaction strength $ \Del $.
For $ \Del = 0 $ all quantum correlations are always zero but for $ \Del \neq 0 $ quantum correlations are generated as time elapses.
From this point we learn that (i) environment plays a constructive role, in contrast to its customary destructive role, inducing quantum correlations (ii) in addition to the interaction of qubits with the environment, the anisotropic part of their own interaction with each other is necessary for creation of quantum correlations.
Generally, to see the possible creation/enhancement of a specific quantum correlation $\mathcal{Q}$, one should compare its steady state value with its initial value. In this way, the condition $ \mathcal{Q}(\rho_{\infty}) > \mathcal{Q}(\rho_0) $ implies the phenomenon of creation/enhancement of the quantum correlation $\mathcal{Q}$.
As top plots of this figure show, generation of QE from an initially separable state, the so-called environment-induced entanglement, requires both interaction between qubits and also their own interaction with the environment; no one is sufficient on its own. 
This constructive role played by the environment in favour of entanglement, in contrast to its customary destructive role, has already been noticed. See for instance \cite{AnWaLu-PA-2007, Hoetala-PRA-2009}. 
%
%
%
Despite the apparent implication from this figure that quantum correlations increase with the strength of interaction $\Del$
, it is important to note that this is not always the case. Consider the steady state as a example. For our parameters, $ \ga = 0.1 \om $ the steady state concurrence $ \mathcal{C}(\infty) = 2 \max \left\{0, \frac{ | \Del | \sqrt{ 4 \om^2 + \ga^2 } - \Del^2} { 4 \Om^2 + \ga^2 } \right\} $ is nonzero only for $ \frac{\Del}{\om} \in (-2.0025, 0 ) \cup (0, 2.0025)  $ and is an increasing function of $ \Del $ only when $ - 2.0025~\om < \Del \leq -0.6188~\om  $ and $ 0 < \Del \leq 0.6188~\om  $.
 
%

\begin{figure} 
\centering
\includegraphics[width=12cm,angle=-0]{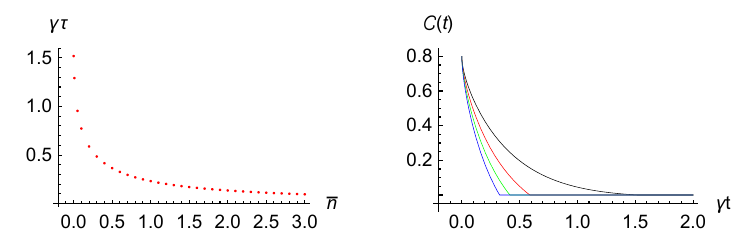}
\caption{
Time of entanglement death, $ \ga \tau $, versus $\bar{n}$ (left panel) and concurrence (right panel) versus 
$\ga t$ for $ \bar{n} = 0 $ (black), $ \bar{n} = 0.2 $ (red), $ \bar{n} = 0.4 $ (green) and $ \bar{n} = 0.6 $ (blue) for $ \Del =  J = 0 $ i.e., qubits solely interact with the common environment not with each other. System is initially described by the mixture \eqref{eq: mix-nonequal-0} with $w = \frac{1}{2}$.
}
\label{fig: pDeaththermal} 
\end{figure}

In the next step we analyse ESD for finite temperatures. This time, the Lindblad equation \eqref{eq: master-temp} must be considered. This equation can still be solved analytically, from which the concurrence for $\Del = J = 0 $ is obtained as follows:
\begin{eqnarray} \label{eq: concur-ther}
\mathcal{C}(t) &=& \max \left\{0, (1-w) e^{-(2\bar{n}+1)\ga t} - \frac{ \sqrt{f(t)} }{ (2\bar{n}+1)^2 } \right\}
\end{eqnarray}
where
\begin{eqnarray}
f(t) &=& a_0(t) + a_1(t) w + a_2(t) w^2 \\
a_0(t) &=& 4 e^{-3 (2\bar{n}+1) \ga t} \sinh^2 \left(\frac{(2\bar{n}+1) \ga t}{2}\right) 
\nonumber \\
&~& \times
\left[ 1 + 4 \bar{n} (\bar{n}+1) e^{(2\bar{n}+1) \ga t/2} \cosh \left(\frac{(2\bar{n}+1) \ga t}{2}\right) \right]^2 \\
a_1(t) &=& 4 (2 \bar{n}+1)^2 e^{- 7 (2\bar{n}+1)\ga t /2 } \sinh \left(\frac{(2\bar{n}+1)\ga t}{2}\right)
\nonumber \\
&~& \times
\left[ 1 + 4 \bar{n} (\bar{n}+1) e^{(2\bar{n}+1)\ga t/2} \cosh \left(\frac{(2\bar{n}+1)\ga t}{2}\right) \right] \\
a_2(t) &=& -2 (2 \bar{n}+1)^4 e^{-3 (2\bar{n}+1)\ga t} \sinh [ (2\bar{n}+1)\ga t ]
\end{eqnarray}
Eq. (\ref{eq: concur-ther}) show that entanglement death occurs here. The time of ESD is obtained from the solution of
\begin{eqnarray} \label{eq: ED-time-ther}
e^{ 2 ( 2\bar{n} + 1 ) \ga t } f(t) &=& ( 2\bar{n} + 1 )^4 ( 1 - w )^2 
\end{eqnarray}
which cannot be solved analytically. Numerical solution of this equation for $ w = \frac{1}{2} $, as a function of $ \bar{n} $, has been given in figure \ref{fig: pDeaththermal}. Right panel of this figure depicts evolution of concurrence for a few values of $ \bar{n} $. ESD is clearly seen being stronger for higher values of $\bar{n}$. This means that at finite temperatures, the temperature speeds up ESD. 
As the left panel of this figure displays, the age of entanglement decreases with $\bar{n}$. 

\begin{figure} 
\centering
\includegraphics[width=12cm,angle=-0]{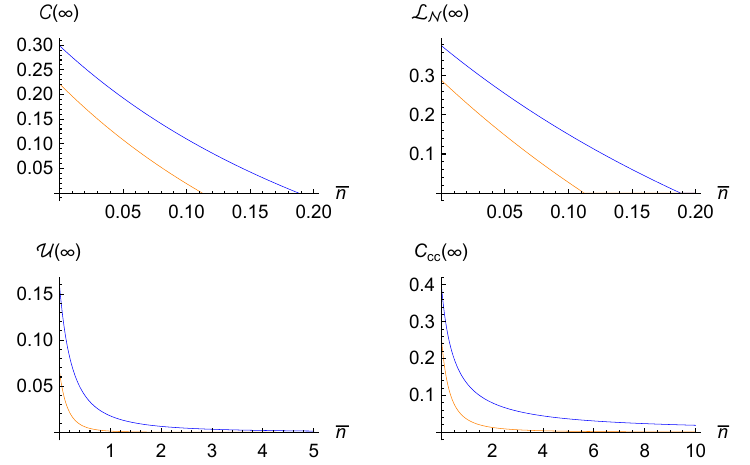}
\caption{
Plots of the finite temperature, steady state quantum correlations, concurrence (left top panel), log negativity (right top panel), LQU (left bottom panel) and CC (right bottom panel), in terms of the average excitation of the thermal bath, $\bar{n}$, for $ \ga = 0.01 \om $ (blue curves) and  $ \ga = 3 \om $ (orange curves) for $ \Del = 0.5 \om $. Note that X-shaped steady state is independent of both the initial state and the isotropic part of qubits' interaction, i.e., the coefficient $J$. 
}
\label{fig: cor-st-temp} 
\end{figure}

Finally, in figure \ref{fig: cor-st-temp} we have plotted the finite temperature, steady state quantum correlations versus the average excitation of the thermal bath, $\bar{n}$, for two values of the emission rate, $ \ga = 0.01 \om $ (magenta curves) and  $ \ga = 3 \om $ (orange curves), for $ \Del = 0.5 \om $. 
As one sees, a steady state level of all quantum correlations is always reached for zero or finite, low temperatures. For nonzero temperatures, all quantum correlations are found to vanish at  a finite temperature. This means none of the quantum correlations are robust against the temperature. However, CC which equals MIN is more robust than others.
Emission rate $\ga$ speeds up the decay of quantum correlations.
%

\section{Summary and conclusions} \label{sec: coclusion}

In this work, our model was a system of two localized qubits interacting through their spins, and also, an external magnetic field was applied. The interaction with the environment was introduced by a non-unitary term in the Lindblad equation. We solved this equation analytically at zero temperature for several X-shaped states and studied the evolution of different quantum correlations for different values of the emission rate $\ga$. After that, it was time to bring the temperature into play and consider the influence of thermal decoherence on quantum correlations. Although the corresponding master equation could be solved analytically, but due to the length of the solution, we considered only the stable solution and examined the dependence of different quantum correlations on temperature in this case.
Our main results can be summarized as follows:
(i) MIN is equal to CC.
(ii) The sudden death of quantum correlations beyond entanglement is not possible, maybe not an unexpected result with respect to their definition.  
(iii) Even if the initial state lacks correlations, the interaction with the environment induces them. In other words, like entanglement, it is possible to induce other quantum correlations by the decoherence.
(iv) In addition to the interaction of qubits with the common environment, their own interaction with each other is necessary to induce quantum correlations.
(v) Although the revival and rebirth of entanglement happens after a finite period of time, the revival of other correlations is much faster.
(vi) In the zero-temperature steady state, LQU captures less quantum correlations than the other quantifiers.
(vii) Temperature speeds up ESD.
(viii) Quantum correlations beyond entanglement are more resistant to the negative effect of temperature to destroy them.
\\
\\
{\bf Acknowledgements:} Support from the University of Qom is acknowledged. 
\\
\\
{\bf Data availability}: This manuscript has no associated data.


\end{document}